\documentclass[letterpaper,twocolumn,10pt]{article}
\usepackage{usenix2024_SOUPS}

\usepackage{tikz}
\usepackage{amsmath}

\usepackage{filecontents}

\usepackage{comment}
\usepackage{booktabs}

\begin{document}

\date{}

\title{\Large \bf Understanding the "Why": \\ Investigating the Design of Privacy Support Tools (PSTs) through the Lens of Psychological Needs}

\def\plainauthor{Hahn, Zimmermann, Althaus, Heinemann, Zimmer, Stöver, Gerber}

\author{
{\rm Sara Hahn$^\star$$^\dagger$$^\ddagger$, 
Verena Zimmermann$^\S$, 
Simon Althaus$^\dagger$$^\ddagger$, 
Andreas Heinemann$^\star$$^\ddagger$, 
Ephraim Zimmer$^\dagger$$^\ddagger$,} \\
{\rm Alina Stöver$^\dagger$$^\ddagger$,  
Nina Gerber$^\dagger$$^\ddagger$}\\
$^\star$Darmstadt University of Applied Sciences \\
$^\dagger$ Technical University of Darmstadt \\
$^\ddagger$ATHENE – National Research Center for Applied Cybersecurity \\
$^\S$ETH Zurich
} 

\maketitle
\thecopyright


\begin{abstract}
Privacy Support Tools (PSTs) help users control their digital data, but face low adoption due to resignation and overwhelm. 
While current regulations aim to enhance privacy control, users still struggle to manage their privacy. 
To address this, we explore how PSTs can be designed for long-term adoption by applying the theoretical lens of psychological needs, as their fulfillment motivates users to use a digital product.
We conducted in-depth interviews with 15 users and identified Security, Autonomy, and Competence as key psychological needs.
From our results, we derive implications and recommendations for the need-sensitive design of PSTs.

\end{abstract}

\section{Introduction}

Remaining control over their private data is one of the crucial challenges in the digital age for users~\cite{Pew2019Americans,acquisti2020secrets,story2021awareness}. 
To facilitate control over private data, EU legislation is currently considering a regulation that specifies requirements for consent management, e.~g., Personal Information Management Systems, which bundle information and management of data collection consent into a single application~\cite{eprivacy,Telekommunikation-Telemedien-Datenschutz-Gesetz}. 
The research literature also presents promising approaches to help users manage their privacy, particularly through so-called Personal Privacy Assistants, 
in alignment with their individual interests~\cite{stover_investigating_2023}. 
In our work, we refer to approaches from both research and legislation and use the term Privacy Support Tools (PSTs) to refer to tools that aid users in protecting their privacy by giving users control over their digital data and obtaining awareness.
While these systems have great potential to make privacy management more accessible for lay users,
users are currently not supported enough by PSTs to remain and regain control over their private data and PSTs currently miss their great potential to empower users in their privacy sovereignty, as users have difficulties adopting them~\cite{story2021awareness}. 
Extensive research has already been conducted on the technical implementation of PSTs (e. g.,~\cite{sadeh2021personalized, almuhimedi2015your, liu2014reconciling}), as well as initial studies capturing the user perspective (e. g.,~\cite{liu_follow_2016, colnago2020PPA, stover_investigating_2023}). 
With our study, we aim to take this a step further by exploring how PSTs should be designed to foster long-term user adoption. 
To inform the design of these PSTs, we use the motivational concept of psychological needs as a theoretical lens.
Psychological needs drive human behavior~\cite{ryan_self-determination_2000} and answer \textit{why} users are using a digital product~\cite{hassenzahl_needs_2010, hassenzahl_experience_2010}.
Their fulfillment motivates users to use a digital product, as research from the field of Human-Computer Interaction shows~\cite{hassenzahl_needs_2010}. 

Therefore, our objective is to understand the privacy-related psychological needs of users and how to consider them in the design of PSTs.
We investigated psychological needs in our in-depth interview study 
with 15 users to understand: \\
\textbf{RQ1} Which psychological needs are relevant for PST users? 
\\
\textbf{RQ2} How can these needs be addressed by PST features? 

We identified \emph{Security}, \emph{Autonomy}, and \emph{Competence} as essential psychological needs for interacting with PSTs. 
The following descriptions are based on Hassenzahl et al.~\cite{hassenzahl_experience_2010}.
\emph{Security} describes feelings of, for example, following pleasant routines. 
\emph{Autonomy} refers to feelings of being in control and acting independently. 
Among other things, feelings of being capable address the need for \emph{Competence}. 
Our main contributions are:
(1) We offer a new perspective on the design of PSTs through the lens of psychological needs. 
(2) Our results provide a deeper understanding of relevant psychological needs that explain why a person is using a product.  
(3) We contribute implications and recommendations for the need-sensitive design of PSTs.

\section{Methodology}

\label{sec:study_1}
\begin{figure}[ht]
  \centering
  \includegraphics[width=0.8\linewidth]{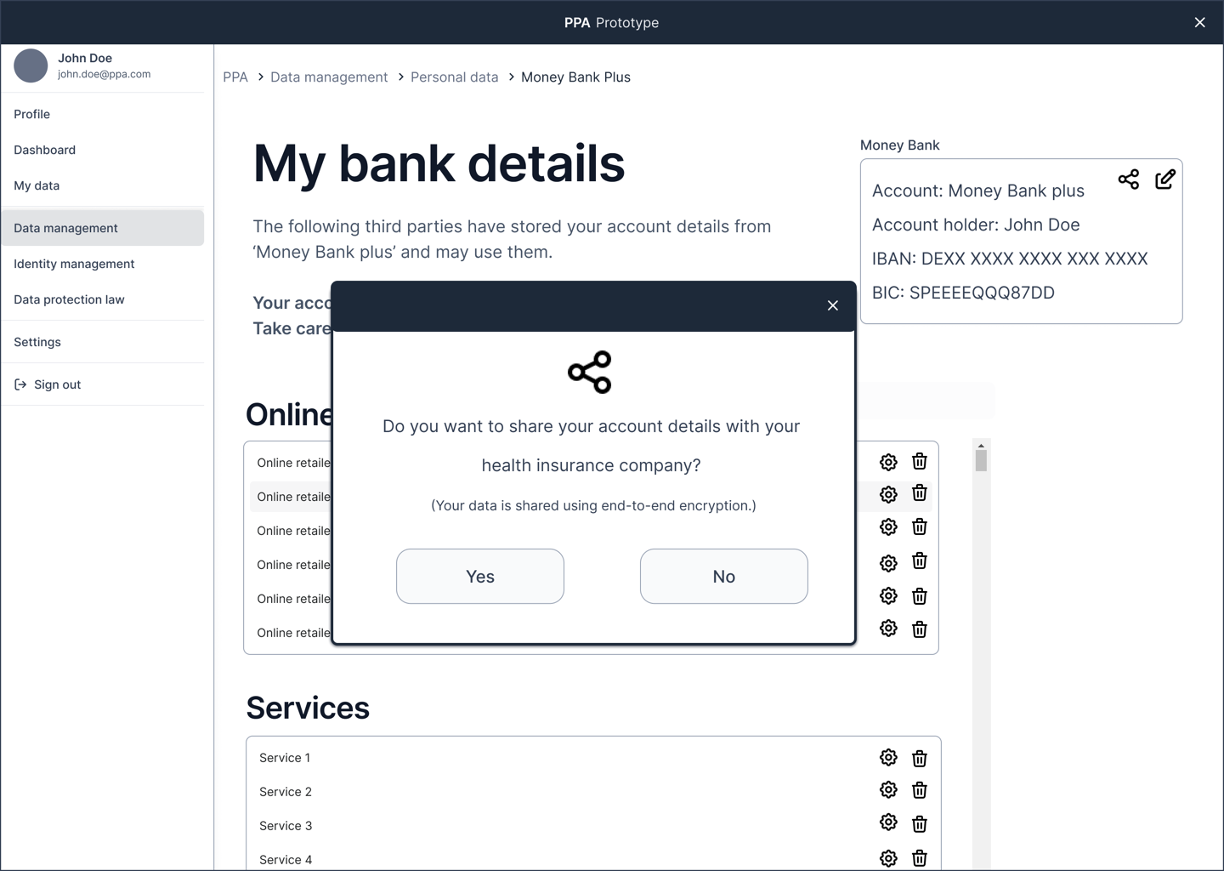}
  \caption{This figure shows a screenshot of a low-fidelity prototype of a PST that served as a basis in the interviews. Here, the user uses the PST to share their bank details with their health insurance company for personal data management.}
  \label{fig:PPA Prototype 1}
\end{figure}

Fifteen participants were recruited from the personal network and university mailing lists.
Most participants were women, between 26 and 30 years old, and university students or in full-time employment.
For compensation, participants from our university could receive course credit.
Our participants reported moderate levels of technical affinity 
and high levels of privacy concerns (captured with the ATI \cite{Franke2019} and IUIPC-8 Scale \cite{gross_validity_2021,Malhotra2004}).
The interview study was conducted remotely via Zoom~\cite{zoom_2023} and consisted of a card-sorting task, followed by an interview and a survey.
The goal of the card-sorting task was to understand which psychological needs are most relevant to users when interacting with PSTs. 
We conducted semi-structured interviews to explore the relevance of psychological needs in specific scenarios, such as deleting and sending personal data to third parties, and to identify design implications. 
The focus was on generating design ideas rather than evaluating the prototype scenarios (see Figure~\ref{fig:PPA Prototype 1}), as this exemplary question shows: "Can you describe another implementation option that
meets [this need]?"
After the interview, participants could adjust their ranking from the initial card-sorting task if they felt it was necessary.
The interview data were analyzed using reflexive thematic analysis informed by Braun and Clarke's approach~\cite{braun_thematic_2012} and processed with MAXQDA \cite{verbi_maxqda_2022}. 
The following themes were identified: 
(1) Relevance of psychological needs in the context of data management and protection;
(2) Interrelations of psychological needs;
Design options that take into account the need for (3) \emph{Security}, 
(4) \emph{Autonomy}, 
(5) \emph{Competence}, 
(6) \emph{Meaning}, 
(7) \emph{Relatedness}, 
(8) \emph{Stimulation}, 
(9) and \emph{Popularity}; and
(10) Other.
Our participants' demographics could have influenced the perceived importance of specific psychological needs.
 Therefore, in future investigations, we aim to recruit representative samples for better transferability of the findings.

All prerequisites for studies with human participants given by our institution's ethics commission were met: We informed all participants about the study conditions and purpose before they were asked to sign the informed consent sheet. 
They could abort the study at any time without negative consequences. 
All participants had the option to have their data deleted at any time by contacting the primary investigator. 
Participation was voluntary. 
They were not exposed to physical or psychological harm exceeding an everyday level.  
The collected data were stored locally and only handled by members of our research group. 
Questionnaire data were captured via SoSci Survey~\cite{leiner_sosci_2019}, which is subject to strict~EU data regulation.

\section{Key Results}
\begin{figure}[ht]
  \centering
  \includegraphics[width=1.0\linewidth]{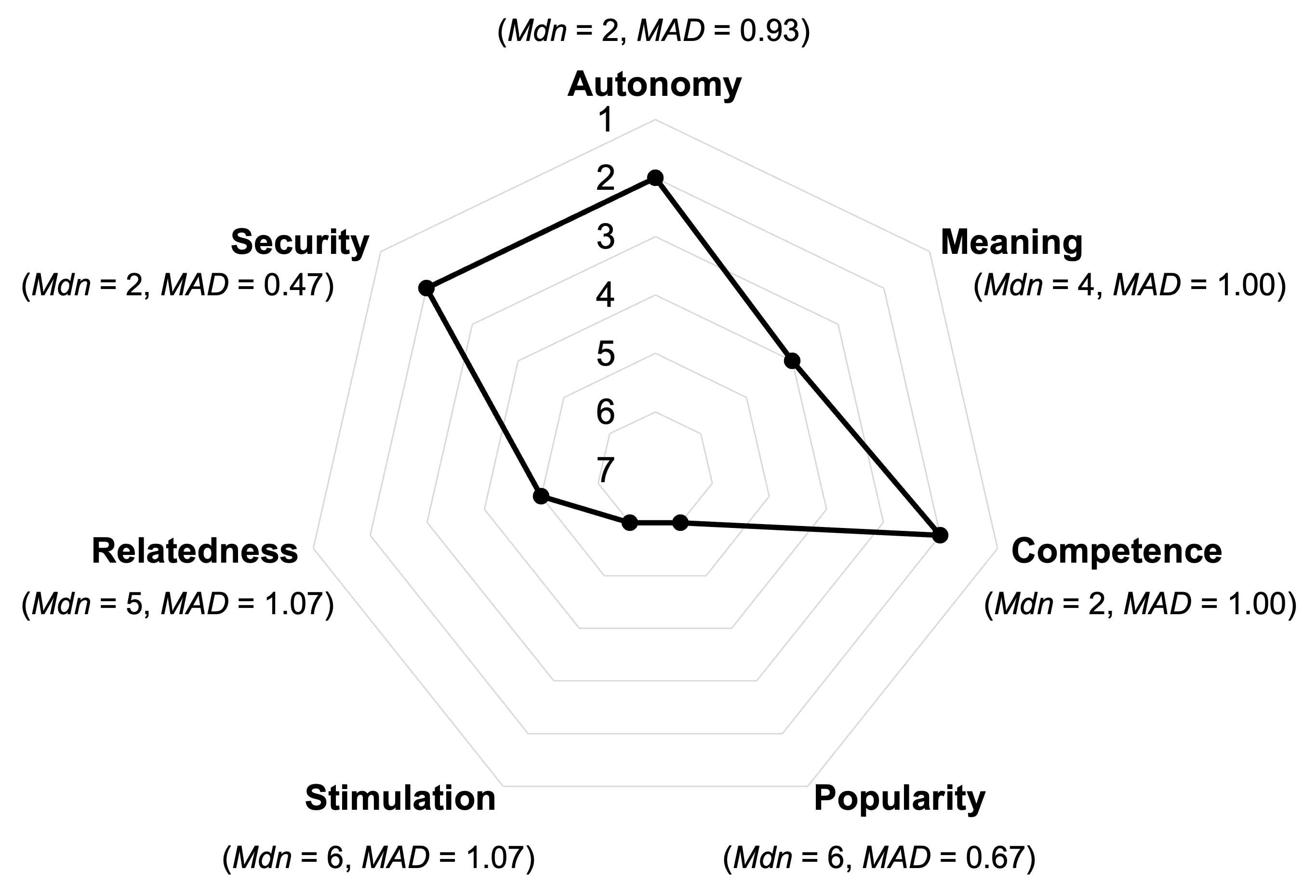}
  \caption{This radar chart shows the determined Median (Mdn) of the relevance ranking (1~=~most relevant to 7~=~least relevant) of the respective psychological need and the Median Absolute Deviation (MAD).}
  \label{fig:Netzdiagramm}
\end{figure}

The needs for \emph{Security}, \emph{Autonomy}, and \emph{Competence} were rated by participants as most relevant for~PSTs, followed by \emph{Meaning} and \emph{Relatedness}. 
The needs rated least relevant were \emph{Stimulation} and \emph{Popularity} (see Figure~\ref{fig:Netzdiagramm}). 
The thematic analysis of the transcripts revealed several possible implementations that address certain psychological needs. 
Our results show that feelings of \emph{Security} counteract perceiving privacy as overwhelming. 
An overview of the data management can fulfill
the need for \emph{Security}, while making settings individually can address the need for \emph{Autonomy}.
According to our participants, feelings of \emph{Autonomy} drive informed privacy-decisions and feelings of in\emph{competence} lead to disengagement. 
To feel \emph{competent}, users wished for a dashboard or feedback concerning their successful protection of their digital data.

\section{Discussion}

We provide novel insights into the psychological needs relevant to PST users, refining prior research by clarifying which needs should be prioritized and how they should be addressed in PST design. 
While existing studies have emphasized the importance of \emph{Autonomy}, \emph{Competence}, and \emph{Security}~\cite{kraus_psychological_2017, ryan_self-determination_2000, zimmermann_why_2019}, our findings add nuance by illustrating concrete ways in which these needs can be supported through design: 
PSTs should align with user expectations and communicate changes transparently (\emph{Security}), allow individual customization (\emph{Autonomy}), and provide feedback mechanisms that enhance the users' sense of control and effectiveness (\emph{Competence}). Unlike prior research on interactive systems, which has often considered \emph{Stimulation} and \emph{Relatedness} as relevant~ \cite{kraus_psychological_2017, zimmermann_why_2019, hassenzahl_needs_2010, gerber_sharing_2017}, our findings suggest that these may not play a significant role in the context of PSTs. 
Some participants even described these needs as contradictory to the fundamental purpose of a PST, emphasizing privacy and individual control over social interaction or novelty. 
Our work highlights the importance of evaluating psychological needs in context rather than assuming their universal applicability -- as their fulfillment is highly context-dependent~\cite{zimmermann_why_2019, hassenzahl_needs_2010} -- and paves the way for the design of need-sensitive PSTs that foster long-term user adoption.

\section*{Acknowledgments}
\textit{Transparency note}: 
This work formed the basis for an additional study and includes verbatim passages from a full paper combining both studies (currently under review at NordiCHI).
This research work was supported 
by the \emph{Deutsche Forschungsgemeinschaft} (DFG, German Research Foundation) -- 251805230/GRK2050, and 
by the National Research Center for Applied Cybersecurity ATHENE. ATHENE is funded jointly by the German Federal Ministry of Research, Technology and Space and the Hessian Ministry of Science and Research, Arts and Culture.

\bibliographystyle{plain}
\bibliography{persuasive_bib_V1.bib}

\appendix
\section*{Appendix}
\section{Prototype}

\section{Results of the Card Sorting Task}

\end{document}